\documentclass[aps,preprint]{revtex4}%
\usepackage{amsfonts}
\usepackage{amsmath}
\usepackage{amssymb}
\usepackage{graphicx}%
\providecommand{\U}[1]{\protect\rule{.1in}{.1in}}

\begin{document}
\title[ ]{Relativistic Hydrogen in Classical Electrodynamics with Classical Zero-point Radiation}
\author{Timothy H. Boyer}
\affiliation{Department of Physics, City College of the City University of New York, New
York, New York 10031}
\keywords{}
\pacs{}

\begin{abstract}
Classical electrodynamics including classical electromagnetic zero-point
radiation leads to a ground state and resonant excited states for a charged
particle in a Coulomb potential. \ These resonant states correspond to integer
values of the action variables analogous to those appearing in the
Bohr-Sommerfeld theory of the hydrogen atom. \ The work on \textit{classical
zero-point radiation} reported here is a continuation of the analysis reported
in 1975, but with the addition of the ideas of \textit{relativity} and
\textit{resonance} between the charge-particle orbit and classical zero-point
radiation. \ 

\end{abstract}
\maketitle

\section{Introduction \ }

\subsection{Old Quantum Theory}

Old quantum theory had its great success with the energy levels of hydrogen.
\ It accepted the ideas of classical particle orbits, but ignored the
associated classical radiation; it simply postulated discrete values of the
action variables. \ In 1924, Born wrote in his \textit{Physics of the Atom},
\textquotedblleft The endeavour to retain the classical mechanics as far as
possible having proved to be a fertile method, we take as our first
requirement that the stationary states of an atomic system shall be
calculated, as far as possible, in accordance with the laws of classical
mechanics, but the classical theory of radiation is
disregarded.\textquotedblright\cite{Born52}\ \ 

Old quantum theory did not give any hint as to why the action variables took
the \textit{integer-times-}$\hbar$ values which were assumed. \ Indeed, the
theory is now treated as a historical precursor because it seemed incapable of
explaining the anomalous Zeeman effect, the energy levels of the hydrogen
molecular ion, and the helium atom. \ Thus the advent of the new quantum
mechanics, introduced by the work of Heisenberg and Schroedinger, was seen as
a great advance. \ Currently, the Schroedinger equation, electron spin, the
Zeeman effect, and the fractional quantum numbers associated with the angular
momentum values of energy levels are treated as \textquotedblleft settled
physics,\textquotedblright\ and contrary ideas are usually dismissed as of
\textquotedblleft no interest\textquotedblright\ or as being completely
erroneous. \ Even Sommerfeld's completely-accurate 1916
calculation\cite{Sommerfeld} of the relativistic corrections for the hydrogen
energy levels is currently seen as \textquotedblleft
incorrect,\textquotedblright\ a mere a \textquotedblleft fortuitous
accident\textquotedblright\ since it contains neither wave mechanics nor
spin.\cite{YM}\cite{Biedenharn} \ 

\subsection{Classical Electrodynamics with Classical Zero-Point Radiation}

On the other hand, for over 60 years, a small group of physicists has explored
a very different point of view, suggesting that at least a significant part of
microscopic physics can be explained as simply classical electromagnetism,
which classical theory includes classical electromagnetic zero-point
radiation. \ In purely classical terms, the theory has successfully explained
Casimir forces, van der Waals forces, low-temperature specific heats of
solids, diamagnetism, superfluid behavior, the blackbody radiation spectrum,
and the absence of atomic collapse.\cite{Brev}\cite{Cole2003} \ In the present
article, we continue this exploration. \ 

A survey article\cite{B1975} in 1975 included the idea of \textit{classical
electromagnetic zero-point radiation}, leading to stability of the hydrogen
ground state. \ However, it did not restrict allowed systems to
\textit{relativistic} systems, and it did not mention \textit{resonance}
between the orbiting charged particle and classical zero-point radiation.
\ Here we suggest that \textit{the integer values for the action variables in
the Bohr-Sommerfeld theory are the result of resonance of the charged particle
orbit with classical electromagnetic zero-point radiation.} \ In related
articles, we will suggest that the electron is a point charged particle
without any spin, and that both the Stern-Gerlach experiment and the Zeeman
effect can be explained by classical electromagnetism which includes classical
zero-point radiation. \ \ 

The restriction to \textit{relativistic} (or approximately relativistic)
systems and the requirement of \textit{resonance} for the orbiting charged
particle in zero-point radiation are ideas completely missing from old quantum
theory. \ The emission of radiation by an accelerating charged particle was
widely accepted when Bohr\cite{Bohr1913} made his proposal in 1913, but the
gain of the right amount of energy on repeated orbital passage through random
classical zero-point radiation does not seem to have been considered. \ 

\subsection{Resonant Phenomena in Classical Zero-Point Radiation}

Classical electromagnetism is quite different from nonrelativistic statistical
mechanics: \textit{classical electromagnetic theory allows} \textit{resonance}
between radiation and matter, as nonrelativistic statistical mechanics does
not. \ For example, in nonrelativistic statistical mechanics, a
one-dimensional charged harmonic oscillator with a natural frequency
$\omega_{0}$ will come to thermal equilibrium at $k_{B}T$ in a bath at
temperature $T$, no matter what value is taken by $\omega_{0}$. \ On the other
hand, in classical electrodynamics, the oscillator will have a dipole
\textit{resonance} with radiation whose frequency matches $\omega_{0}$ at the
position of the oscillator. \ If the oscillator has finite spatial extent,
then there can be resonance at the multiples of $\omega\,_{0}$%
.\cite{SHOarticle} \ \ 

The old Bohr picture as a model of an atom is easy to imagine; it is often
repeated in textbooks of modern physics\cite{ModernTexts} despite the claim
that it is outmoded by newer developments. \ Our suggestion is that the model,
with integer values for the action variables, is valid for resonance because
of radiation energy balance with classical electromagnetic zero-point
radiation. \ \textit{The resonant orbit is merely a radiation--energy-balanced
orbit involving a classical particle orbit but with an integer values for all
the action variables because resonance requires the repeated accumulation of
effects.} \ The integer values come from the Lorentz-invariant zero-point
radiation. \ For one-electron systems such as the hydrogen atoms, this makes
the situation particularly simple. \ For resonance, one chooses a classical
particle trajectory having integer values for all the action variables. \ This
works for the Bohr-Wilson-Sommerfeld nonrelativistic atom and for the
Sommerfeld relativistic atom. \ Further corrections (such as the Lamb shift)
involve more detailed radiation balance. \ \ 

\subsection{Coulomb Potential\ }

Classical zero-point radiation is Lorentz invariant, and the Coulomb potential
is part of \textit{relativistic} classical electrodynamics. \ The ground state
involves resonance for all the integer multiples of the oscillator
ground-state frequency. \ The resonances above the ground state in general
involve the \textit{dipole} interaction between radiation and the charged
particle. \ These excited state resonances are unstable and will decay to the
ground state with the emission of electromagnetic radiation corresponding to
the Bohr frequency condition. \ In the limit where the speed of light is
regarded as very large compared to the speed of the charged particle, the
\textit{results} (but not the analyses) become those of nonrelativistic
Bohr-Wilson-Sommerfeld theory. \ 

\section{Relativistic Charged Particle in a Coulomb Potential}

\subsection{Mechanical System Hamiltonian}

Within classical electromagnetic theory, the hydrogen atom consists of a very
massive nucleus and a (relativistic) electron moving in a classical orbit
determined by the classical electrostatic interaction of the nucleus, and the
driving of the classical zero-point radiation. \ The motion is assumed in the
$xy$-plane, where the electrostatic force from the nucleus provides the
centripetal acceleration%
\begin{equation}
M_{0}\frac{d}{dt}\left(  \frac{\mathbf{v}}{\sqrt{1-\left(  v/c\right)  ^{2}}%
}\right)  =-\widehat{r}\frac{e^{2}}{r^{2}}.
\end{equation}
In spherical coordinates, the Hamiltonian\cite{Biedenharn2} for this system is
given by its kinetic energy and Coulomb potential energy%
\begin{equation}
H=\sqrt{\left(  p_{r}^{2}+p_{\phi}^{2}\right)  c^{2}+\left(  M_{0}%
c^{2}\right)  ^{2}}-\frac{e^{2}}{r}, \label{Ham}%
\end{equation}
where $\mathbf{p=}\widehat{r}p_{r}+\widehat{\phi}p_{\phi}=M_{0}\gamma
\mathbf{v}$, $M_{0}$ is the rest mass, $\mathbf{v}\,$\ is the velocity, and
$\gamma=\left[  1-\left(  v/c\right)  ^{2}\right]  ^{-1/2}$. \ The orbital
angular momentum is given by%
\begin{equation}
L_{\phi}=rp_{\phi}=M_{0}\gamma r^{2}\dot{\phi}.
\end{equation}
Both the energy $U_{e}$ and angular momentum $L_{\phi}$ of the system are
constants of the classical motion. \ 

\subsection{Orbit and Energy}

We wish to solve for the orbit and the particle energy. \ In the traditional
fashion, we introduce
\begin{equation}
s=1/r\text{ \ }%
\end{equation}
and note
\begin{equation}
p_{r}=M_{0}\gamma\dot{r}=M_{0}\gamma\frac{dr}{d\phi}\frac{d\phi}{dt}=\left(
M_{0}\gamma r^{2}\frac{d\phi}{dt}\right)  \left(  \frac{1}{r^{2}}\frac
{dr}{d\phi}\right)  =-L_{\phi}\left(  ds/d\phi\right)  .
\end{equation}
Inserting the variables $s$ and $L_{\phi}$ into Eq. (\ref{Ham}), we have%
\begin{equation}
\left(  \frac{H+e^{2}s}{M_{0}c^{2}}\right)  ^{2}=1+\left[  \left(
\frac{L_{\phi}c}{e^{2}}\right)  \left(  \frac{e^{2}}{M_{0}c^{2}}\right)
\right]  ^{2}\left[  \left(  \frac{ds}{d\phi}\right)  ^{2}+s^{2}\right]
\end{equation}
Differentiating this equation with respect to $\phi$ and then cancelling
common factors,$\,\ $we obtain%
\begin{equation}
\frac{d^{2}s}{d\phi^{2}}+\Gamma^{2}\left(  s-\frac{e^{2}H}{\Gamma^{2}L_{\phi
}^{2}c^{2}}\right)  =0, \label{SH}%
\end{equation}
where
\begin{equation}
\Gamma=\sqrt{1-\left(  \frac{e^{2}}{L_{\phi}c}\right)  ^{2}}, \label{Gamma}%
\end{equation}
and%
\begin{equation}
\frac{L_{\phi}c}{e^{2}}\Gamma=\sqrt{\left(  \frac{L_{\phi}c}{e^{2}}\right)
^{2}-1}.
\end{equation}

Equation (\ref{SH}) has the solution%
\begin{equation}
s=\zeta\cos\left(  \Gamma\phi+\beta\right)  +\frac{e^{2}H}{\Gamma^{2}L_{\phi
}^{2}c^{2}},
\end{equation}
where $\zeta$ and $\beta$ are constants of integration. \ Choosing the
perihelion as occurring at $\phi=0,$ the constant $\beta$ vanishes and the
equation takes the simpler form
\begin{equation}
\frac{1}{r}=\zeta\cos\left(  \Gamma\phi\right)  +\frac{e^{2}H}{\Gamma
^{2}L_{\phi}^{2}c^{2}}=\zeta\cos\left(  \Gamma\phi\right)  +\left(  \frac
{U}{M_{0}c^{2}}\right)  \left(  \frac{e^{2}}{\Gamma L_{\phi}c}\right)
^{2}\left(  \frac{M_{0}c^{2}}{e^{2}}\right)  ,
\end{equation}
or the familiar rosette form%
\begin{equation}
r=\frac{a(1-\epsilon^{2})}{1+\epsilon\cos\left(  \Gamma\phi\right)  },
\end{equation}
where at perihelion $r_{per}=a\left(  1-\epsilon\right)  $ and at aphelion
$r_{aph}=a\left(  1+\epsilon\right)  $ so that $2a=r_{per}+r_{aph}$. \ Then
the constant energy is given in terms of the constant angular momentum at
perihelion or aphelion as
\begin{equation}
U=\sqrt{\left(  L_{\phi}/r_{per}\right)  ^{2}c^{2}+\left(  M_{0}c^{2}\right)
^{2}}-\frac{e^{2}}{r_{per}}=\sqrt{\left(  L_{\phi}/r_{aph}\right)  ^{2}%
c^{2}+\left(  M_{0}c^{2}\right)  ^{2}}-\frac{e^{2}}{r_{aph}}. \label{Utest}%
\end{equation}
The orbital equation gives
\begin{equation}
\zeta=\frac{\epsilon}{a\left(  1-\epsilon^{2}\right)  }\text{ }%
\end{equation}
and%
\begin{equation}
\text{\ }\frac{e^{2}H}{L_{\phi}^{2}c^{2}\Gamma^{2}}=\frac{1}{a\left(
1-\epsilon^{2}\right)  }. \label{e2H}%
\end{equation}

\subsection{Action-Angle Variables}

The orbital angles and momenta for an orbit in the $xy$-plane can be
reexpressed in terms of the action-angle variables $\phi_{r},\phi_{\phi}%
,J_{r},J_{\phi}$. \ The angular momentum is a constant giving
\begin{equation}
J_{\phi}=\frac{1}{2\pi}\oint p_{\phi}d\phi=L_{\phi}. \label{Jphi}%
\end{equation}
The radial variable is
\begin{align}
J_{r}  &  =\frac{1}{2\pi}\oint p_{r}dr=\frac{1}{2\pi}\int_{0}^{2\pi\Gamma
}\left[  M_{0}\gamma\left(  \frac{dr}{d\phi}\right)  \dot{\phi}\right]
\left(  \frac{dr}{d\phi}d\phi\right) \nonumber\\
&  =\frac{1}{2\pi}\int_{0}^{2\pi\Gamma}L_{\phi}\left(  \frac{1}{r}\frac
{dr}{d\phi}\right)  ^{2}d\phi=L_{\phi}\Gamma\left[  \frac{1}{2\pi}\int%
_{0}^{2\pi}\frac{\epsilon^{2}\sin^{2}\psi}{(1+\epsilon\cos\psi)^{2}}%
d\psi\right] \nonumber\\
&  =L_{\phi}\Gamma\left[  \frac{1}{\sqrt{1-\epsilon^{2}}}-1\right]  ,
\label{JrJphi}%
\end{align}
since the integral is the same as the nonrelativistic integral and involves
only the parameters of an ellipse. \ \ 

\subsection{Energy, Frequency, and Eccentricity in Terms of Action-Angle
Variables}

Now using Eqs. (\ref{Gamma}), (\ref{Utest}), (\ref{e2H}), and (\ref{JrJphi})
we obtain the relativistic classical Hamiltonian[Goldstein498] in terms of the
action variables of the charged particle system%
\begin{equation}
U\left(  J_{r},J_{\theta},J_{\phi}\right)  =M_{0}c^{2}\left(  1+\left[
\frac{e^{2}/c}{J_{r}+\sqrt{\left(  J_{\theta}+J_{\phi}\right)  ^{2}-\left(
e^{2}/c\right)  ^{2}}}\right]  ^{2}\right)  ^{-1/2}.\label{UJr}%
\end{equation}
It follows that the radial frequency $\omega_{r}$ is%
\begin{equation}
\frac{\partial U}{\partial J_{r}}=\omega_{r}=\frac{M_{0}c^{2}\left(
e^{2}/c\right)  ^{2}}{\left[  \left(  e^{2}/c\right)  ^{2}+\left(  J_{r}%
+\sqrt{\left(  J_{\theta}+J_{\phi}\right)  ^{2}-\left(  e^{2}/c\right)  ^{2}%
}\right)  ^{2}\right]  ^{3/2}},
\end{equation}
and the frequencies $\omega_{\theta}~$and $\omega_{\phi}$ are%
\begin{equation}
\omega_{\theta}=\omega_{\phi}=\frac{M_{0}c^{2}\left(  e^{2}/c\right)  ^{2}%
}{\sqrt{1-\left[  e^{2}/\left[  \left(  J_{\theta}+J_{\phi}\right)  c\right]
^{2}\right]  }\left[  \left(  e^{2}/c\right)  ^{2}+\left(  J_{r}+\sqrt{\left(
J_{\theta}+J_{\phi}\right)  ^{2}-\left(  e^{2}/c\right)  ^{2}}\right)
^{2}\right]  ^{3/2}}.\label{wtheta}%
\end{equation}
Thus two distinct frequencies are involved giving%
\begin{equation}
\omega_{r}=\Gamma\omega_{\theta}=\Gamma\omega_{\phi}\text{ \ \ where
\ \ }\Gamma=\sqrt{1-\left\{  e^{2}/\left[  \left(  J_{\theta}+J_{\phi}\right)
c\right]  \right\}  ^{2}}.\label{wr}%
\end{equation}
The semimajor axis is
\begin{equation}
a=\frac{e^{2}U}{\left(  M_{0}c^{2}\right)  ^{2}-U^{2}},\label{aU}%
\end{equation}
the eccentricity is%
\begin{equation}
\epsilon=\sqrt{1-\frac{1}{\left[  1+J_{r}/\sqrt{\left(  J_{\theta}+J_{\phi
}\right)  ^{2}-\left(  e^{2}/c\right)  ^{2}}\right]  ^{2}}}=\sqrt{1-\frac
{1}{\left\{  1+J_{r}/\left[  \left(  J_{\theta}+J_{\phi}\right)
\Gamma\right]  \right\}  ^{2}}},\label{epsilon}%
\end{equation}
and the time is connected to the orbital motion by%
\begin{equation}
t=\int_{0}^{\phi}d\phi^{\prime}\frac{r^{2}\left[  U+e^{2}/r\right]  }%
{c^{2}\left(  J_{\theta}+J_{\phi}\right)  }=\left[  1+\frac{\left(
1-\epsilon^{2}\right)  }{\left[  \left(  M_{0}c^{2}/U\right)  ^{2}-1\right]
}\right]  ^{-1/2}\times\int_{0}^{\phi}\frac{d\phi^{\prime}}{c}\left(
\frac{r^{2}U}{e^{2}}+r\right)  .
\end{equation}

\subsection{Circular Orbit}

For a \textit{circular} orbit in the $xy$-plane, the angular momentum is
\begin{equation}
J_{\phi}=M_{0}\gamma rv=\frac{M_{0}}{\sqrt{1-\left(  v/c\right)  ^{2}}}rv,
\end{equation}
while the orbital angle $\phi$ advances uniformly in time $t,$%
\begin{equation}
\phi_{\phi}\left(  t\right)  =\omega_{\phi}t+\phi_{\phi0},
\end{equation}
with the angle $\phi_{r}$ becoming
\begin{equation}
\phi_{r}\left(  t\right)  =\omega_{r}t+\phi_{r0}=\Gamma\omega_{\phi}%
t+\phi_{r0}.
\end{equation}

From Eq. (\ref{UJr})The energy in a circular orbit is%

\begin{equation}
U=\frac{M_{0}c^{2}}{\gamma}=M_{0}c^{2}\sqrt{1-\left(  \frac{e^{2}}{J_{\phi}%
c}\right)  ^{2}},
\end{equation}
giving a ratio%
\begin{equation}
\frac{J_{\phi}}{U}=\frac{J_{\phi}}{M_{0}c^{2}\sqrt{1-\left[  e^{2}/\left(
J_{\phi}c\right)  \right]  ^{2}}}.
\end{equation}
The orbital frequency $\omega_{\phi}$ takes the form%

\begin{align}
\omega_{\phi}  &  =\left(  \frac{\partial H}{\partial J_{\phi}}\right)
=\frac{M_{0}c^{2}}{\sqrt{1-\left[  e^{2}/\left(  J_{\phi}c\right)  \right]
^{2}}}\left(  \frac{e^{2}}{c}\right)  ^{2}\frac{1}{J_{\phi}^{3}}\nonumber\\
&  =\left(  \frac{M_{0}c^{3}}{e^{2}}\right)  \frac{1}{\sqrt{1-\left[
e^{2}/\left(  J_{\phi}c\right)  \right]  ^{2}}}\left(  \frac{e^{2}}{J_{\phi}%
c}\right)  ^{3}=\left(  \frac{M_{0}\gamma c^{3}}{e^{2}}\right)  \left(
\frac{e^{2}}{J_{\phi}c}\right)  ^{3} \label{wphiH}%
\end{align}
while the radius is%
\begin{align}
r  &  =\frac{v}{\omega}=\frac{e^{2}}{\omega J_{\phi}}=\frac{e^{2}}{J_{\phi}%
}\frac{e^{2}\sqrt{1-\left[  e^{2}/\left(  J_{\phi}c\right)  \right]  ^{2}}%
}{M_{0}c^{3}}\left(  \frac{J_{\phi}c}{e^{2}}\right)  ^{3}\nonumber\\
&  =\frac{e^{2}}{M_{0}c^{2}}\sqrt{1-\left[  e^{2}/\left(  J_{\phi}c\right)
\right]  ^{2}}\left(  \frac{J_{\phi}c}{e^{2}}\right)  ^{2}=\frac{e^{2}}%
{M_{0}\gamma c^{2}}\left(  \frac{J_{\phi}c}{e^{2}}\right)  ^{2},
\end{align}
and the speed is therefore%
\begin{equation}
\omega r=v=\frac{e^{2}}{J_{\phi}}.
\end{equation}
We note that in a circular orbit $J_{r}=0,$ while the ratio
\begin{equation}
\frac{J_{\phi}\omega_{\phi}}{U}=\frac{1}{1-\left[  e^{2}/\left(  J_{\phi
}c\right)  \right]  ^{2}}\left(  \frac{e^{2}}{J_{\phi}c}\right)  ^{2}=\frac
{1}{\left[  J_{\phi}c/\left(  e^{2}\right)  \right]  ^{2}-1}=\left[  \left(
\frac{J_{\phi}c}{e^{2}}\right)  ^{2}-1\right]  ^{-1},
\end{equation}
which is a function of the total orbital angular momentum $J_{\phi}$ with
$e^{2}/c<J_{\phi}$.\cite{incompat} \ \ 

\section{Interaction with Classical Electromagnetic Zero-Point Radiation}

\subsection{Need for Classical Electromagnetic Zero-Point Radiation}

Our relativistic classical mechanical analysis for a charged particle in a
Coulomb (or Kepler) potential has reached the point where we need to know the
values of the action variables in order to describe the motion of the system.
\ However, classical mechanics contains no fundamental constants, except
Cavendish's constant connected to gravity. \ Our analysis does not involve
gravity, and hence there is no fundamental value with the mechanical motion
associated with the action variables. \ Thus, we must go outside of
relativistic classical \textit{mechanics} in order to describe nature within
classical theory. \ 

If we prefer to retain classical theory, we turn to classical
\textit{electrodynamics, }since many of the systems involve interactions
between charged particles. \ In order to account for the spatial dependence of
the observed force between uncharged conducting parallel plates (the Casimir
effect\cite{Casimir}\cite{Casimir2}), we must assume that nature includes
random classical electromagnetic zero-point radiation with a Lorentz-invariant
spectrum.\cite{Marshall} \ The assumption of Lorentz-invariance fixes the
spectrum of classical zero-point radiation up to an undetermined overall
constant. \ The spatial dependence of the force between the plates fits with
classical zero-point radiation. \ However, in order to account for the
\textit{magnitude} of the observed force, we must choose the scale of the
classical zero-point radiation as a number proportional to Planck's constant
$\hbar$; the average zero-point radiation energy $U^{zp}\left(  \omega\right)
$ per normal mode of frequency $\omega$ must be
\begin{equation}
U^{zp}\left(  \omega\right)  =\left(  1/2\right)  \hbar\omega. \label{ZPR}%
\end{equation}
But this random classical radiation cannot be limited to just one phenomenon;
it will force all charged particles into random motion with a scale associated
with $\hbar.$ \ This classical electromagnetic zero-point radiation will
indeed give a scale to the action variables $J_{i}$\ of a charged particle $e$
in a Coulomb potential. \ 

\subsection{Integer Values Associated with Radiation}

Although the scale of zero-point radiation may fix the ground state values for
the particle action variables, the association with radiation implies an
integer connection with higher multipole behavior. \ If classical zero-point
radiation is Lorentz-invariant, then it must be isotropic and follow the
irreducible representations of the rotation group. \ Since the
\textit{irreducible presentations} \textit{of a group} are associated with
integer values, we expect integer values to be associated with the action
variables of the charged particles. \ Also, periodic motion will allow the
possibility of resonances between orbiting charged particles and the driving
radiation. \ The possibility of \textit{resonance} associated with radiation
is a crucial idea which is missing from old quantum theory and indeed from
many previous analyses. \ 

\subsection{Classical Electromagnetic Radiation}

The source-free classical zero-point radiation in a very large
\textit{spherical} cavity of radius $\mathsf{R}$ can be written
as\cite{Jackson}%

\begin{align}
\mathbf{E}(\mathbf{r,}t)  &  =\operatorname{Re}\sum\nolimits_{n=1}^{\infty
}\sum\nolimits_{l=1}^{\infty}\sum\nolimits_{m=-l}^{m=l}\left\{  \exp\left[
i\left(  -k_{nl}^{M}ct+\theta_{nlm}^{M}\right)  \right]  \left[  ia_{nlm}%
^{M}j_{l}\left(  k_{nlm}^{M}r\right)  \mathbf{X}_{l,m}\left(  \theta
,\phi\right)  \right]  \right. \nonumber\\
&  \left.  +\exp\left[  i\left(  -k_{nlm}^{E}ct+\theta_{nlm}^{E}\right)
\right]  \left[  a_{nlm}^{E}/\left(  -ik_{nlm}^{E}\right)  \right]
\nabla\times\left[  j_{l}\left(  k_{nlm}^{E}r\right)  \mathbf{X}_{lm}\left(
\theta,\phi\right)  \right]  \right\}  , \label{Ezprt}%
\end{align}
and%

\begin{align}
\mathbf{B}(\mathbf{r,}t)  &  =\operatorname{Re}\sum\nolimits_{n=1}^{\infty
}\sum\nolimits_{l=1}^{\infty}\sum\nolimits_{m=-l}^{m=l}\left\{  \exp\left[
i\left(  -k_{nlm}^{E}ct+\theta_{nlm}^{E}\right)  \right]  \left[  ia_{nlm}%
^{E}j_{l}\left(  k_{nlm}^{E}r\right)  \mathbf{X}_{l,m}\left(  \theta
,\phi\right)  \right]  \right. \nonumber\\
&  \left.  +\exp\left[  i\left(  -k_{nlm}^{M}ct+\theta_{nlm}^{M}\right)
\right]  \left[  a_{nlm}^{M}/\left(  ik_{nlm}^{M}\right)  \right]
\nabla\times\left[  j_{l}\left(  k_{nlm}^{M}r\right)  \mathbf{X}_{lm}\left(
\theta,\phi\right)  \right]  \right\}  , \label{Bzprt}%
\end{align}
where $a_{nlm}^{E}$ and $a_{nlm}^{M}$ are the amplitudes of the electric and
magnetic modes, $j_{l}$ is the spherical Bessel function of order $l$,
$\mathbf{X}_{l,m}\left(  \theta,\phi\right)  $ is the vector spherical
harmonic, and $\theta_{nlm}^{E},\theta_{nlm}^{E}$\ are the random phases
distributed uniformly on $(0,2\pi]$ and independently for each mode.\cite{EH}%
\cite{Rice} \ The amplitudes of the spherical standing waves correspond to the
spectrum $U^{zp}\left(  c\mathbf{k}\right)  =\frac{1}{2}\hbar ck$ given in Eq.
(\ref{ZPR}), where the angular frequency $\omega=ck=c\left\vert \mathbf{k}%
\right\vert $, and the randomness is introduced by the random phases for the
electric and magnetic modes $\theta_{nlm}^{E}$ and $\theta_{nlm}^{M}$. \ The
representation of the rotation group refers to the integers $l$ and $m$ which
are unrelated to the frequency $\omega$ of the normal mode.

Here we have the vector spherical harmonic\cite{Jackson2}%
\begin{equation}
\mathbf{X}_{lm}\left(  \theta,\phi\right)  =\frac{\mathbf{L}Y_{lm}\left(
\theta,\phi\right)  }{\sqrt{l\left(  l+1\right)  }}=\frac{\mathbf{r\times
\nabla}Y_{l,m}\left(  \theta,\phi\right)  }{i\sqrt{l\left(  l+1\right)  }},
\end{equation}
where the spherical harmonics are given by\cite{Jackson3}%
\begin{align}
Y_{lm}\left(  \theta=\pi/2,\phi\right)   &  =\left[  \sqrt{\frac{\left(
2l+1\right)  }{4\pi}\frac{\left(  l-m\right)  !}{\left(  l+m\right)  !}}%
P_{l}^{m}\left(  \cos\theta\right)  \right]  _{\theta=\pi/2}\exp\left[
im\phi\right] \nonumber\\
&  =Y_{lm}\left(  \pi/2,0\right)  \left\{  \cos\left[  m\phi\right]
+i\sin\left[  m\phi\right]  \right\}  ,
\end{align}
and the only complex part is the exponential behavior in $\phi$, $\exp\left[
im\phi\right]  $. \ 

For random radiation, we have the scale given by\cite{Disguised}%

\begin{equation}
\left\vert a_{nlm}^{E}\right\vert ^{2}=\frac{16\pi\left(  k_{nl}^{E}\right)
^{2}}{\mathsf{R}}\left[  U_{nlm}\left(  k_{nlm}^{E}\right)  \right]  .
\label{aEnlm2}%
\end{equation}

The situation for \textit{magnetic} modes $a_{nlm}^{M}$ is exactly analogous.
\ Magnetic radiation modes contribute for a charged particle in a circular
orbit beginning at $l=2$ with a magnitude comparable to the \textit{electric}
multipole of order $l=3$.

For the limit of large radius $\mathsf{R}$ of the enclosing sphere containing
standing waves, we have\cite{Disguised} $dk=\pi dn/\mathsf{R}$ and \
\begin{equation}
\sum\nolimits_{n=1}^{\infty}\rightarrow\int_{0}^{\infty}dn=\int_{0}^{\infty
}dk\frac{\mathsf{R}}{\pi}=\int_{0}^{\infty}d\omega\frac{\mathsf{R}}{\pi c}.\,
\label{SumC}%
\end{equation}
If we use random phases for closely-space radiation modes labeled by $nlm$,
then the amplitude involves the average energy $U_{nlm}.$ \ The number of
normal modes per unit (angular) frequency per unit volume is $\omega
^{2}/\left(  \pi^{2}c^{3}\right)  $ which gives an energy per unit angular
frequency interval per unit volume $\left[  \omega^{2}/\left(  \pi^{2}%
c^{3}\right)  \right]  U\left(  \omega\right)  .$ \ 

\section{Gain of Energy for a Charge in a Circular Orbit in Zero-Point
Radiation}

\subsection{Equation of Motion for the Charge Driven by Random Radiation}

A charge $e$ in a \textit{circular} orbit in the $xy$-plane will experience a
force due to the $\phi$-component of the electric field $E_{\phi}^{E}%
(r_{e},\pi/2,\phi_{e}\left(  t\right)  ,t)$ at the angular location $\phi
_{e}\left(  t\right)  $ of the charge. \ Newton's second law gives%
\begin{equation}
Mr_{e}^{2}\frac{d^{2}\phi_{e}\left(  t\right)  }{dt^{2}}=r_{e}eE_{\phi}%
^{E}(\mathbf{r}_{e}\left(  t\right)  ,t), \label{Mr2dphi}%
\end{equation}
where $M=M_{0}\gamma$ is the relativistic particle mass which is assumed not
to change during the short time $\tau$. \ For the solution of this equation,
we use variation of parameters analogous to Born's work\cite{Born2} for the
linear harmonic oscillator. \ Now the \textit{source-free} solutions of the
homogeneous equation $Mr_{0}^{2}\ddot{\phi}=0$ for the circular orbit are
$\phi\left(  t\right)  =a$ and $\phi\left(  t\right)  =bt$ where $a$ and $b$
are constants. \ Then it follows that the particular solution $\phi_{p}(t)$ of
Eq. (\ref{Mr2dphi}) is obtained by variation of parameters as\cite{Greenberg}%

\begin{align}
Mr_{e}\phi_{p}(t)  &  =-\int_{0}^{t^{\prime}=t}dt^{\prime}\left[  eE_{\phi
}^{E}(\mathbf{r}_{e}\left(  t^{\prime}\right)  ,t^{\prime})\right]  t^{\prime
}+\int_{0}^{t^{\prime}=t}dt^{\prime}\left[  eE_{\phi}^{E}(\mathbf{r}%
_{e}\left(  t^{\prime}\right)  ,t^{\prime})\right]  t\nonumber\\
&  =\int_{0}^{t^{\prime}=t}dt^{\prime}\left[  eE_{\phi}^{E}(\mathbf{r}%
_{e}\left(  t^{\prime}\right)  ,t^{\prime})\right]  \left(  t-t^{\prime
}\right)  . \label{Mrphip}%
\end{align}
Note that the angular velocity $d\phi_{p}\left(  t\right)  /dt$ follows as%
\begin{align}
Mr_{e}\frac{d\phi_{p}(t)}{dt}  &  =-\left[  eE_{\phi}^{E}(\mathbf{r}%
_{e}\left(  t\right)  ,t)\right]  t+\left[  eE_{\phi}^{E}(r_{e},\pi/2,\phi
_{e}\left(  t\right)  ,t)\right]  t\nonumber\\
&  +\int_{0}^{t^{\prime}=t}dt^{\prime}\left[  eE_{\phi}^{E}(\mathbf{r}%
_{e}\left(  t^{\prime}\right)  ,t^{\prime})\right] \nonumber\\
&  =\int_{0}^{t^{\prime}=t}dt^{\prime}\left[  eE_{\phi}^{E}(\mathbf{r}%
_{e}\left(  t^{\prime}\right)  ,t^{\prime})\right]  , \label{Mr0v}%
\end{align}
and indeed the equation of motion (\ref{Mr2dphi}) is satisfied, since when we
take the second derivative with respect to time $t$, we obtain \
\begin{equation}
Mr_{e}\frac{d^{2}\phi_{p}(t)}{dt^{2}}=\left[  eE_{\phi}^{E}(\mathbf{r}%
_{e}\left(  t\right)  ,t)\right]  .
\end{equation}
Thus the proposed solution in Eq. (\ref{Mrphip}) indeed satisfies the
differential equation (\ref{Mr2dphi}), and also satisfies the boundary
conditions $\phi_{p}\left(  0\right)  =0$ and $\dot{\phi}_{p}\left(  0\right)
=0$ for the circular orbit. \ 

\subsection{Energy Gain from Random Radiation}

The energy delivered by the random radiation during a short time interval
$\tau$ involving many oscillations but little change in the mechanical angular
momentum $J_{\phi}$ of the particle is
\begin{align}
W\left(  \tau\right)   &  =\int_{0}^{\tau}dt\left[  eE_{\phi}^{E}%
(\mathbf{r}_{e}\left(  t\right)  ,t)\right]  \left[  r_{e}\frac{d\phi_{p}%
(t)}{dt}\right] \nonumber\\
&  =\frac{1}{M}\int_{0}^{\tau}dt\left[  eE_{\phi}^{E}(\mathbf{r}_{e}\left(
t\right)  ,t)\right]  \left\{  \int_{0}^{t^{\prime}=t}dt^{\prime}\left[
eE_{\phi}^{E}(\mathbf{r}_{e}\left(  t^{\prime}\right)  ,t^{\prime})\right]
\right\} \nonumber\\
&  =\frac{e^{2}}{M}\int_{0}^{\tau}dt\int_{0}^{t^{\prime}=t}dt^{\prime}\left[
E_{\phi}^{E}(\mathbf{r}_{e}\left(  t\right)  ,t)\right]  \left[  E_{\phi}%
^{E}(\mathbf{r}_{e}\left(  t^{\prime}\right)  ,t^{\prime})\right]  ,
\label{WPtauP}%
\end{align}
where again $M=M_{0}\gamma$ is the relativistic mass. \ The integrand in Eq.
(\ref{WPtauP}) is \textit{symmetric} under interchange of $t$ and $t^{\prime}%
$. \ \ But then we can carry through Born's symmetrizing. \ Thus, the double
integral over the isosceles triangular region for $t$ and $t^{\prime}$ in Eq.
(\ref{WPtauP}) is the same as obtained by integrating first in $t$ from
$t^{\prime}$ to $\tau$, and then in $t^{\prime}$ from $0$ to $\tau$ as%

\begin{align}
&  \int_{0}^{\tau}dt\int_{0}^{t^{\prime}=t}dt^{\prime}\left[  E_{\phi}%
^{E}(\mathbf{r}_{e}\left(  t\right)  ,t)\right]  \left[  E_{\phi}%
^{E}(\mathbf{r}_{e}\left(  t^{\prime}\right)  ,t^{\prime})\right] \nonumber\\
&  =\int_{0}^{\tau}dt^{\prime}\int_{t=t^{\prime}}^{t=\tau}dt\left[  E_{\phi
}^{E}(\mathbf{r}_{e}\left(  t\right)  ,t)\right]  \left[  E_{\phi}%
^{E}(\mathbf{r}_{e}\left(  t^{\prime}\right)  ,t^{\prime})\right]  .
\end{align}
However, we may interchange the prime and unprime labels and add half the
expressions to obtain%
\begin{equation}
W\left(  \tau\right)  =\frac{e^{2}}{2M}\int_{0}^{t=\tau}dt\int_{0}^{t^{\prime
}=\tau}dt^{\prime}\left[  E_{\phi}^{E}(\mathbf{r}_{e}\left(  t\right)
,t)\right]  \left[  E_{\phi}^{E}(\mathbf{r}_{e}\left(  t^{\prime}\right)
,t^{\prime})\right]  , \label{Wtaue2}%
\end{equation}
where now both time integrals are from $0$ to $\tau$. \ 

\subsection{Averaging Over the Random Phases of the Zero-Point Radiation}

Now we need the \textit{average} energy absorbed in the short time $\tau$
involving many passages over the charged-particle orbit. \ Thus, equation
(\ref{Wtaue2}) becomes%
\begin{equation}
\left\langle W\left(  \tau\right)  \right\rangle _{\theta^{E}}=\frac{e^{2}%
}{2M}\int_{0}^{\tau}dt\int_{0}^{\tau}dt^{\prime}\left\langle \left[  E_{\phi
}^{E}(\mathbf{r}_{e}\left(  t\right)  ,t)\right]  \left[  E_{\phi}%
^{E}(\mathbf{r}_{e}\left(  t^{\prime}\right)  ,t^{\prime})\right]
\right\rangle _{\theta^{E}}, \label{AvWtau}%
\end{equation}
where we average over the random phases $\theta^{E}$ of the driving radiation. \ 

The $\mathbf{r}_{e}\left(  t\right)  $ refers to the periodic motion of the
charged particle in the Coulomb potential and has nothing to do with the
phases of the driving zero-point radiation. \ However, we must maintain the
$\mathbf{r}_{e}\left(  t\right)  $ dependence which corresponds to periodic
particle motion in time. \ Then the $\phi$-component of the electric field due
to zero-point radiation can be rewritten as
\begin{align}
\left[  E_{\phi}^{E}(\mathbf{r}_{e}\left(  t\right)  ,t)\right]   &
=\operatorname{Re}E_{\phi}^{E}\left(  \mathbf{r}_{e}\left(  t\right)
,0\right)  \exp\left[  -i\omega^{E}t+i\theta^{E}\right] \nonumber\\
&  =E_{\phi}^{E}\left(  \mathbf{r}_{e}\left(  t\right)  ,0\right)  \cos\left[
\omega^{E}t-\theta^{E}\right]  .
\end{align}

Random radiation can be treated by introducing random phases for the waves.
\ However, there is already a phase in the problem, that of the charged
particle in its orbit. \ Thus we must be careful to preserve the randomness of
the radiation waves relative to the phase already present in the orbital
motion. \ In equation (\ref{AvWtau}), there are two phases corresponding to
the two different radiation waves at times $t$ and $t^{\prime}$ involved in
the integrals. \ However, if the two phases are different, the integrals over
many periods will give vanishing results. \ Thus the only contribution which
survives the integration is that where a radiation wave is matched with
itself. \ However, this radiation wave will still retain its randomness
compared with the phase of the orbiting charge. \ 

We are interested in averaging over the random phases $\theta_{n^{\prime
}l^{\prime}m^{\prime}}^{E}$ to obtain $\left\langle W\left(  \tau\right)
\right\rangle _{\theta^{E^{\prime}}}$ as in Eq. (\ref{AvWtau}). \ We require%
\begin{equation}
\left\langle \cos\left[  \theta_{nlm}^{E}\right]  \cos\left[  \theta
_{n^{\prime}l^{\prime}m^{\prime}}^{E}\right]  \right\rangle _{\theta
^{E^{\prime}}}=\left\langle \sin\left[  \theta_{nlm}^{E}\right]  \sin\left[
\theta_{n^{\prime}l^{\prime}m^{\prime}}^{E}\right]  \right\rangle
_{\theta^{E^{\prime}}}=\frac{1}{2}\delta_{nlm,n^{\prime}l^{\prime}m^{\prime}},
\end{equation}
and%
\begin{equation}
\left\langle \cos\left[  \theta_{nlm}^{E}\right]  \sin\left[  \theta
_{n^{\prime}l^{\prime}m^{\prime}}^{E}\right]  \right\rangle _{\theta
^{E^{\prime}}}=0.
\end{equation}

\ Then for the electric multipole radiation, averaging and then summing to
eliminate the $\delta_{nlm,n^{\prime}l^{\prime}m^{\prime}}$, we have%
\begin{align}
&  \left\langle \left[  E_{\phi}^{E}(\mathbf{r}_{e}\left(  t\right)
,t)\right]  \left[  E_{\phi}^{E}(\mathbf{r}_{e}\left(  t^{\prime}\right)
,t^{\prime})\right]  \right\rangle _{\theta^{E^{\prime}}}\nonumber\\
&  =\left\langle \sum\nolimits_{nlm}E_{\phi}^{E}\left(  \mathbf{r}_{e}\left(
t\right)  ,0\right)  \cos\left[  \omega^{E}t-\theta_{nlm}^{E}\right]
\sum\nolimits_{n^{\prime}l^{\prime}m^{\prime}}E_{\phi}^{E}\left(
\mathbf{r}_{e}\left(  t^{\prime}\right)  ,0\right)  \cos\left[  \omega
^{E}t^{\prime}-\theta_{n^{\prime}l^{\prime}m^{\prime}}^{E}\right]
\right\rangle _{\theta^{E^{\prime}}}\nonumber\\
&  =\sum\nolimits_{nlm}E_{nlm-\phi}^{E}(\mathbf{r}_{e}\left(  t\right)
,0)\sum\nolimits_{n^{\prime}l^{\prime}m^{\prime}}E_{n^{\prime}l^{\prime
}m^{\prime}-\phi}^{E}(\mathbf{r}_{e}\left(  t^{\prime}\right)  ,0)\cos\left[
\omega_{n^{\prime}l^{\prime}m^{\prime}}^{E}\left(  t-t^{\prime}\right)
\right]  \frac{1}{2}\delta_{nlm,n^{\prime}l^{\prime}m^{\prime}}\nonumber\\
&  =\frac{1}{2}\sum\nolimits_{nlm}\left[  E_{nlm-\phi}^{E}(\mathbf{r}%
_{e}\left(  t\right)  ,0)\right]  \left[  E_{nlm-\phi}^{E}(\mathbf{r}%
_{e}\left(  t^{\prime}\right)  ,0)\right]  \cos\left[  \omega_{nlm}^{E}\left(
t-t^{\prime}\right)  \right]  .
\end{align}
Then the average energy absorbed in Eq. (\ref{AvWtau}) is
\begin{align}
&  \left\langle W^{E}\left(  \tau\right)  \right\rangle _{\theta^{E^{\prime}}%
}\nonumber\\
&  =\frac{e^{2}}{2M}\int_{0}^{\tau}dt\int_{0}^{\tau}dt^{\prime}\left\langle
\left[  E_{\phi}^{E}(\mathbf{r}_{e}\left(  t\right)  ,t)\right]  \left[
E_{\phi}^{E}(\mathbf{r}_{e}\left(  t^{\prime}\right)  ,t^{\prime})\right]
\right\rangle \nonumber\\
&  =\frac{e^{2}}{2M}\int_{0}^{\tau}dt\int_{0}^{\tau}dt^{\prime}\frac{1}{2}%
\sum\nolimits_{nlm}\left[  E_{nlm-\phi}^{E}(\mathbf{r}_{e}\left(  t\right)
,0)\right]  \left[  E_{nlm-\phi}^{E}(\mathbf{r}_{e}\left(  t^{\prime}\right)
,0)\right]  \cos\left[  \omega_{nlm}^{E}\left(  t-t^{\prime}\right)  \right]
. \label{wtauq}%
\end{align}
Now we can use the expansion
\begin{equation}
\cos\left[  \omega_{nlm}^{E}\left(  t-t^{\prime}\right)  -\theta_{nlm}%
^{E}\right]  =\cos\left[  \omega_{nlm}^{E}t-\theta_{nlm}^{E}\right]
\cos\left[  \omega_{nlm}^{E}t^{\prime}\right]  +\sin\left[  \omega_{nlm}%
^{E}t-\theta_{nlm}^{E}\right]  \sin\left[  \omega_{nlm}^{E}t^{\prime}\right]
,
\end{equation}
and therefore can rewrite Eq. (\ref{wtauq}) as%
\begin{align}
\left\langle W^{E}\left(  \tau\right)  \right\rangle  &  =\frac{e^{2}}{4M}%
\sum\nolimits_{nlm}\left\{  \left[  \int_{0}^{\tau}dt\left[  E_{nlm-\phi}%
^{E}(\mathbf{r}_{e}\left(  t\right)  ,0)\right]  \cos\left[  \omega_{nlm}%
^{E}t\right]  \right]  ^{2}\right. \nonumber\\
&  +\left.  \left[  \int_{0}^{\tau}dt\left[  E_{nlm-\phi}^{E}(\mathbf{r}%
_{e}\left(  t\right)  ,0)\right]  \sin\left[  \omega_{nlm}^{E}t\right]
\right]  ^{2}\right\}  \label{avWtaue2}%
\end{align}

Now the charged particle is going around with frequency $\omega_{e-n}$ in a
circular orbit of radius $r_{e-n}$, so that its speed is $\omega_{e-n}%
r_{e-n}=v_{e-n}$. \ Accordingly, the first integral in the \textit{electric}
multipole field in Eq. (\ref{avWtaue2}) requires the expansion%

\begin{equation}
E_{\phi}^{E}\left[  \mathbf{r}_{e}\left(  t\right)  ,0\right]  =\sum
\nolimits_{n^{\prime}lm}^{\infty}E_{n^{\prime}lm-\phi}^{E}\left(
k_{n^{\prime}lm}^{E}r_{e}\right)  \cos\left[  \omega_{e-n}t\right]  .
\end{equation}
Then including the time behavior of the electric field, we have%
\begin{align}
&  \int_{0}^{\tau}dt\left[  E_{n^{\prime}lm-\phi}^{E}(\mathbf{r}_{e}\left(
t\right)  ,0)\right]  \cos\left[  \omega_{n^{\prime}lm}^{E}t\right]
\nonumber\\
&  =\int_{0}^{\tau}dt\left[  \sum\nolimits_{n^{\prime}lm}^{\infty}%
E_{n^{\prime}lm-\phi}^{E}\left(  k_{n^{\prime}lm}^{E}r\right)  \cos\left[
\omega_{e-n}t\right]  \right]  \cos\left[  \omega_{n^{\prime}lm}^{E}t\right]
\nonumber\\
&  =\int_{0}^{\tau}dt\sum\nolimits_{n^{\prime}lm}^{\infty}E_{n^{\prime}%
lm-\phi}^{E}\left(  k_{n^{\prime}lm}^{E}r\right)  \frac{1}{2}\left\{
\cos\left[  \left(  \omega_{e-n}-\omega_{nlm}^{E}\right)  t\right]
+\cos\left[  \left(  \omega_{e-n}+\omega_{nlm}^{E}\right)  t\right]  \right\}
\nonumber\\
&  =\sum\nolimits_{n^{\prime}lm}^{\infty}E_{n^{\prime}lm-\phi}^{E}\left(
k_{n^{\prime}lm}^{E}r\right)  \frac{1}{2}\left\{  \frac{\sin\left[  \left(
\omega_{e-n}-\omega_{n^{\prime}lm}^{E}\right)  \tau\right]  }{\left(
\omega_{e-n}-\omega_{n^{\prime}lm}^{E}\right)  }+\frac{\sin\left[  \left(
\omega_{e-n}+\omega_{n^{\prime}lm}^{E}\right)  \tau\right]  }{\left(
\omega_{e-n}+\omega_{n^{\prime}lm}^{E}\right)  }\right\}  ,
\end{align}
where $\omega_{e-n}$ is the frequency of the charged particle in its orbit.
\ For a circular orbit, the expression $\left[  E_{nlm-\phi}^{E}(r_{e}%
,\pi/2,0,0)\right]  $ does not involve time $t$ and so can be taken outside
the time integral. \ 

For a very large spherical enclosure, the sum turns into an integral, and we
may use Eqs. (\ref{Ezprt}) and (\ref{aEnlm2}) giving
\begin{align}
&  \left\langle W_{l,m}^{E}\left(  \tau\right)  \right\rangle \nonumber\\
&  =\frac{e^{2}}{4M}\sum\nolimits_{nlm}\left[  E_{nlm-\phi}^{E}(r_{e}%
,\pi/2,0,0)\right]  ^{2}\left\{  \left[  \int_{0}^{\tau}dt\cos\left[  \left(
\omega_{e}-\omega_{n^{\prime}l^{\prime}m^{\prime}}^{E}\right)  t\right]
\right]  ^{2}\right. \nonumber\\
&  +\left.  \left[  \int_{0}^{\tau}dt\sin\left[  \left(  \omega_{e}%
-\omega_{n^{\prime}l^{\prime}m^{\prime}}^{E}\right)  t\right]  \right]
^{2}\right\} \nonumber\\
&  =\frac{e^{2}}{4M}\left(  \int_{0}^{\infty}d\omega^{E}\frac{\mathsf{R}}{\pi
c}\right)  \left[  m^{2}\left[  \frac{1}{x}\left(  \frac{d\left[
xj_{l}\left(  x\right)  \right]  }{d\left(  x\right)  }\right)  \right]
_{x=mv/c}Y_{lm}\frac{a_{lm}^{E}}{\sqrt{l\left(  l+1\right)  }}\right]
_{\theta=\pi/2,\phi=0}^{2}\nonumber\\
&  \times\left\{  \left[  \int_{0}^{\tau}dt\cos\left[  \left(  \omega
_{e}-\omega^{E}\right)  t\right]  \right]  ^{2}+\left[  \int_{0}^{\tau}%
dt\sin\left[  \left(  \omega_{e}-\omega^{E}\right)  t\right]  \right]
^{2}\right\} \nonumber\\
&  =\frac{e^{2}}{4M}\left(  \int_{0}^{\infty}d\omega^{E}\frac{\mathsf{R}}{\pi
c}\right)  m^{4}\left[  \frac{1}{x}\left(  \frac{d\left(  xj_{l}\left(
x\right)  \right)  }{dx}\right)  \right]  _{x=mv/c}^{2}\frac{\left\vert
Y_{lm}\right\vert ^{2}}{l\left(  l+1\right)  }\left[  \frac{16\pi\left(
\omega_{nl}^{E}\right)  ^{3}}{c^{2}\mathsf{R}}\left\langle J_{rad}\left(
\omega_{nl}^{E}\right)  \right\rangle \right] \nonumber\\
&  \times\left\{  \left[  \int_{0}^{\tau}dt\cos\left[  \left(  \omega
_{e}-\omega^{E}\right)  t\right]  \right]  ^{2}+\left[  \int_{0}^{\tau}%
dt\sin\left[  \left(  \omega_{e}-\omega^{E}\right)  t\right]  \right]
^{2}\right\} \nonumber\\
&  =\frac{4\pi e^{2}}{M}\left(  \int_{0}^{\infty}d\omega^{E}\frac{\mathsf{R}%
}{\pi c}\right)  m^{4}\left[  \frac{1}{x}\left(  \frac{d\left(  xj_{l}\left(
x\right)  \right)  }{dx}\right)  \right]  _{x=mv/c}^{2}\frac{\left\vert
Y_{lm}\right\vert ^{2}}{l\left(  l+1\right)  }\left[  \frac{\left(
\omega_{nl}^{E}\right)  ^{3}}{c^{2}\mathsf{R}}\left\langle J_{rad}\left(
\omega_{nl}^{E}\right)  \right\rangle \right] \nonumber\\
&  \times\left\{  \left[  \frac{\sin^{2}\left[  \left(  \omega_{e}-\omega
^{E}\right)  \tau\right]  }{\left(  \omega_{e}-\omega^{E}\right)  ^{2}%
}\right]  +\left[  \frac{\left(  1-\cos\left[  \left(  \omega_{e}-\omega
^{E}\right)  \tau\right]  \right)  ^{2}}{\left(  \omega_{e}-\omega^{E}\right)
^{2}}\right]  \right\}  `
\end{align}

For a single spherical wave mode $l,m$ at frequency $\omega^{E}$, we integrate
over $\omega^{E}$ as
\begin{equation}
\int_{0}^{\infty}d\omega^{E}\left[  \frac{\sin^{2}\left[  \left(  \omega
_{e}-\omega^{E}\right)  \tau\right]  }{\left(  \omega_{e}-\omega^{E}\right)
^{2}}\right]  \approxeq\int_{-\infty}^{\infty}d\omega^{E}\left[  \frac
{\sin^{2}\left[  \left(  \omega_{e}-\omega^{E}\right)  \tau\right]  }{\left(
\omega_{e}-\omega^{E}\right)  ^{2}}\right]  =\pi\tau,
\end{equation}
and%
\begin{equation}
\int_{0}^{\infty}d\omega^{E}\left[  \frac{\left(  1-\cos\left[  \left(
\omega_{e}-\omega^{E}\right)  \tau\right]  \right)  ^{2}}{\left(  \omega
_{e}-\omega^{E}\right)  ^{2}}\right]  \approxeq\int_{-\infty}^{\infty}%
d\omega^{E}\left[  \frac{\left(  1-\cos\left[  \left(  \omega_{e}-\omega
^{E}\right)  \tau\right]  \right)  ^{2}}{\left(  \omega_{e}-\omega^{E}\right)
^{2}}\right]  =\pi\tau.
\end{equation}

For the ground state where $\omega^{E}=\omega_{e}$, the integral in
$\omega^{E}$ collapses because of the resonance at $\omega^{E}=\omega_{e},$
and we have
\begin{equation}
P_{l,m}^{gain-E}=\frac{e^{2}}{M_{0}\gamma_{e}}\frac{\left(  \omega^{E}\right)
^{3}}{c^{3}}\left\langle \left[  J_{rad}\left(  \omega_{nl}^{E}\right)
\right]  \right\rangle \left\{  8\pi\frac{m^{4}\left[  Y_{lm}\right]  ^{2}%
}{l\left(  l+1\right)  }\left[  \frac{1}{x}\frac{d\left[  xj_{l}(x)\right]
}{dx}\right]  _{x=mv/c}^{2}\right\}  .
\end{equation}
This expression is to be compared to the power lost in the same radiation mode
given by Burko\cite{Burko}%
\begin{align}
&  P_{l,m}^{loss-E}=8\pi\frac{e^{2}}{c^{3}}m^{4}\omega_{e}^{4}r_{e}^{2}%
\frac{l\left(  l+1\right)  }{\left(  2l+1\right)  ^{2}}\left[  Y_{l,m}\left(
\pi/2,0\right)  \right]  ^{2}\left[  \frac{1}{l+1}j_{l+1}\left(  m\omega
_{e}r_{e}/c\right)  -\frac{1}{l}j_{l-1}\left(  m\omega_{e}r_{e}/c\right)
\right]  ^{2}\nonumber\\
&  =M_{0}c^{2}\left(  \frac{M_{0}c^{3}}{e^{2}}\right)  \frac{\left[
e^{2}/\left(  J_{\phi}c\right)  \right]  ^{8}}{1-\left[  e^{2}/\left(
J_{\phi}c\right)  \right]  ^{2}}\left\{  8\pi\frac{m^{4}\left[  Y_{l,m}\left(
\pi/2,0\right)  \right]  ^{2}}{l\left(  l+1\right)  }\left[  \frac{1}{x}%
\frac{d\left[  xj_{l}(x)\right]  }{dx}\right]  _{x=mv/c}^{2}\right\}
\end{align}

We obtain the condition for energy balance by equating the power lost and
gained,
\begin{equation}
P_{l,m}^{loss-E}=P_{l,m}^{gain-E}.
\end{equation}
\ We remove the common factors of
\begin{equation}
8\pi\frac{m^{4}\left\vert Y_{lm}\right\vert ^{2}}{l\left(  l+1\right)
}\left[  \frac{1}{x}\frac{d\left[  xj_{l}(mx)\right]  }{dx}\right]
_{x=v/c}^{2}, \label{Common}%
\end{equation}
and simplify using Eq. (\ref{wphiH}) to find
\begin{equation}
J_{\phi}=J_{rad}=\hbar. \label{JeJrad}%
\end{equation}
In the ground state, the connection in Eq. (\ref{JeJrad}) gives energy balance
between the loss of energy due to emission of radiation as dipole radiation
and the gain of energy from the dipole driving by zero-point radiation for the
relativistic charge. \ 

We expect that the ground state is completely stable, so that the energy
balance should hold not just for the dipole radiation, but for every radiation
mode. \ Burko\cite{Burko} gives the radiation emission for all modes when a
charged particle is moving in a circular orbit. \ \ We notice that the
variation from one mode to the next is contained in the factors which were
common to both the loss and gain of energy by the orbiting charge and are
given in the display (\ref{Common}). \ We expect this situation to continue
for the magnetic radiation modes.

\section{Resonant Excited States}

\subsection{Circular Resonant Excited States Have Lower Frequency}

The \textit{resonant excited states} of a charged particle in a Coulomb
potential are expected to be unstable and to decay with radiation emission
going down in energy to the stable ground state. \ Thus we will consider only
the energy balance for \textit{dipole} radiation in the resonant excited
states. \ It is the absence of equilibrium for the higher multipole which
leads to the unstable behavior of the resonant excited states. \ On radiation
decay, it is usually the dipole radiation which is emitted as the charged
particle changes orbits from one excited state to a different state. \ 

For the circular resonant excited state labeled by $n$, the orbital angular
momentum $J_{e-n}$ is larger than that for the ground state $J_{e-1}.$
\begin{equation}
J_{e-n}=nJ_{e-1},
\end{equation}
as is the radius for the circular orbit
\begin{equation}
r_{e-n}=\left(  \frac{e^{2}}{M_{0}\gamma_{e-n}c^{2}}\right)  \left(
\frac{J_{e-n}c}{e^{2}}\right)  ^{2}=\left(  \frac{e^{2}}{M_{0}c^{2}}\right)
\sqrt{1-\left(  \frac{e^{2}}{J_{e-n}c}\right)  ^{2}}\left(  \frac{J_{e-n}%
c}{e^{2}}\right)  ^{2},
\end{equation}
while the orbital angular frequency is smaller%
\begin{equation}
\omega_{e-n}=\left(  \frac{M_{0}\gamma_{e-n}c^{3}}{e^{2}}\right)  \left(
\frac{e^{2}}{J_{e-n}c}\right)  ^{3}=\left(  \frac{M_{0}c^{3}}{e^{2}}\right)
\frac{1}{\sqrt{1-\left[  e^{2}/\left(  J_{e-n}c\right)  \right]  ^{2}}}\left(
\frac{e^{2}}{J_{e-n}c}\right)  ^{3}.
\end{equation}
These equations agree with those given for a circular orbit with the square
roots arising from the relativistic mass expression. \ %

\begin{equation}
P_{l,m}^{gain-E}=\frac{e^{2}}{M_{0}\gamma_{e}}\frac{\left(  \omega^{E}\right)
^{3}}{c^{3}}\left[  J_{rad}\left(  \omega_{nl}^{E}\right)  \right]  \left\{
8\pi\frac{m^{4}\left[  Y_{lm}\right]  ^{2}}{l\left(  l+1\right)  }\left[
\frac{1}{x}\frac{d\left[  xj_{l}(x)\right]  }{dx}\right]  _{x=mv/c}%
^{2}\right\}  \label{Pgain}%
\end{equation}%
\begin{align}
&  P_{l,m}^{loss-E}\nonumber\\
&  =M_{0}c^{2}\left(  \frac{M_{0}c^{3}}{e^{2}}\right)  \frac{\left[
e^{2}/\left(  J_{\phi}c\right)  \right]  ^{8}}{1-\left[  e^{2}/\left(
J_{\phi}c\right)  \right]  ^{2}}\left\{  8\pi\frac{m^{4}\left[  Y_{l,m}\left(
\pi/2,0\right)  \right]  ^{2}}{l\left(  l+1\right)  }\left[  \frac{1}{x}%
\frac{d\left[  xj_{l}(x)\right]  }{dx}\right]  _{x=mv/c}^{2}\right\}
\label{Ploss}%
\end{align}

Removing the common factors given in curly brackets in Eqs. (\ref{Pgain}) and
(\ref{Ploss}), the requirement for energy balance in a circular orbit is that%
\begin{equation}
\frac{e^{2}}{M_{0}\gamma_{e-n}}\frac{\left(  \omega_{e-1}/n^{3}\right)  ^{3}%
}{c^{3}}\left\langle \left[  J_{rad}\left(  \omega_{nl}^{E}\right)  \right]
\right\rangle =M_{0}c^{2}\left(  \frac{M_{0}c^{3}}{e^{2}}\right)
\frac{\left[  e^{2}/\left(  nJ_{e-1}c\right)  \right]  ^{8}}{1-\left[
e^{2}/\left(  nJ_{e-1}c\right)  \right]  ^{2}}.
\end{equation}
Simplifying, we find the energy-balance requirement is%
\begin{equation}
\frac{1}{nJ_{e-1}}J_{rad}=1. \label{1dnh}%
\end{equation}
The relationship implies that the zero-point radiation should be delivering to
the charge $n$ times as much power as being lost in the radiation emission.

\subsection{Higher Radiation Multipoles and Repeated Forces}

Because the orbit of the charge has finite extent, $r_{e-n}\neq0,$ it cannot
be treated as a point dipole. \ The charge density of the charge is given by
\begin{equation}
\rho_{e}(r,\phi,0,t)=e\delta^{2}\left[  \mathbf{r-}\widehat{x}r\cos\left(
\omega t\right)  -\widehat{y}r\sin\left(  \omega t\right)  \right]  ,
\end{equation}
which is a periodic function of time $t,$ but involves all the
\textit{multiples} of the fundamental orbital frequency\cite{Jackson4} \ Thus,
the charge density is given by
\begin{equation}
\rho_{e}(r,\phi,0,t)=e\sum\nolimits_{m=0}^{\infty}\rho_{m}(r)\cos\left[
m\omega_{e}t\right]
\end{equation}
where the Fourier transform is
\begin{equation}
\rho_{m}\left(  r\right)  =\frac{4\pi}{\omega_{e}}\int_{0}^{\omega_{e}/\left(
2\pi\right)  }dt\,\rho_{e}(r,\phi,0,t)\cos\left[  m\omega_{e}t\right]  .
\end{equation}

The zero-point radiation mode at frequency of the ground state, $\omega
_{rad}=\omega_{e-1},$ is going around faster that the slower orbital motion
$\omega_{e-n}=\omega_{e-1}/n^{3}$ at larger radius. \ Thus the rotating
radiation wave sweeps over the orbiting charged particle $n$ times, because
the zero-point radiation is still pushing the charged particle with the same
force and at the same frequency $\omega_{rad}=\omega_{n-1}=\omega_{e-1}$.
\ The charged particle's orbital frequency is much lower than when it is in
the ground state. \ For example, if $n=2,$ the speed of the charge in its
orbit is only half that of the ground state,%
\begin{equation}
r_{e-2}\omega_{e-2}=\left(  r_{e-1}n^{2}\right)  \left(  \omega_{e-1}%
/n^{3}\right)  =v_{e-2}=e^{2}/\left(  2J_{e-1}\right)  =e^{2}/\left(
2\hbar\right)  . \label{rewwe2}%
\end{equation}
This means that the wave will pass over the charge $n$ times while the charge
goes once around its orbit. \ But if the radiation wave now passes over the
particle $n$ times while the charge goes around its orbit once, then it picks
up $n$ times as much energy. \ This is exactly the factor of $n$ needed to
give energy balance for the \textit{dipole} radiation in the resonant excited
state in Eq. (\ref{1dnh}). \ 

\section{Nonrelativistic Limit}

In the limit as the speed of the charged particle is regarded as very small
compared to the speed of light $c,$ all the equations above go over to their
nonrelativistic limits. \ It seem fascinating that all the nonrelativistic
formulae of the Bohr theory are finite and do not involve the constant $c$.
\ Thus as $c$ is regraded as very large compared to any particle speeds, the
formulae of the relativistic expressions go over to the nonrelativistic
mechanical expressions where the constant $c$ does not enter. \ For example,
the particle speed (as always) is simply
\begin{equation}
v_{e-n}=\frac{e^{2}}{n\hbar},
\end{equation}
the nonrelativistic orbital radius is%
\begin{equation}
r_{e-n}=\frac{e^{2}}{M_{0}c^{2}}\left[  \sqrt{1-\left(  \frac{e^{2}}{J_{\phi
}c}\right)  ^{2}}\left(  \frac{J_{\phi}c}{e^{2}}\right)  ^{2}\right]
\rightarrow\frac{n^{2}\hbar^{2}}{M_{0}e^{2}},
\end{equation}
and the orbital frequency is
\begin{equation}
\omega_{e-n}=\frac{M_{0}c^{3}}{e^{2}}\left[  \frac{1}{\sqrt{1-\left[
e^{2}/\left(  J_{\phi}c\right)  \right]  ^{2}}}\left(  \frac{e^{2}}{J_{\phi}%
c}\right)  ^{3}\right]  \rightarrow\frac{M_{0}e^{4}}{n^{3}\hbar}.
\end{equation}

Radiation emission and absorption \textit{does} indeed involve $c.$ \ However,
radiation energy balance in zero-point radiation is independent of the actual
value for $c$, and the balance is hidden. \ Indeed, the Bohr theory postulates
that a charged particle in certain preferred orbits simply does \textit{not}
radiate. \ 

\section{Closing Summary}

In 1913 when Bohr suggested that in certain preferred orbits an electron did
not radiate, it occasioned profound skepticism among physicists. \ The
successes of classical electrodynamics were well known, and the emission of
radiation by an accelerating charged particle was widely accepted. \ However,
despite its unsettling ideas, Bohr's theory gave the correct results for the
wavelengths of the line spectra from hydrogen, even giving the correct reduced
mass correction for the positive helium ion. \ In 1916,
Sommerfeld\cite{Sommerfeld}extended Bohr's ideas to give the correct fine
structure of the hydrogen spectral lines. \ 

Today, the situation in physics is vastly different. \ Classical
electromagnetism is regarded as holding for macroscopic situations, but
quantum mechanics with its unusual ideas holds sway for the microscopic
domain. \ And as Feynman has claimed, \textquotedblleft I think I can safely
say that nobody understands quantum mechanics.\textquotedblright

In the present article, we suggest that Bohr's ideas of 1913 can be understood
in terms of classical electromagnetism with random classical zero-point
radiation. \ Although the idea of a \textit{classical zero-point radiation}
was considered by Nernst\cite{Nernst} in 1916, it never gained traction in the
physics community. \ The idea was treated seriously and extensively for the
harmonic oscillator by Marshall\cite{Marshall2} in 1963, and later by others.
\ But free fields or linear potentials gave the only successes. \ 

A summary article\cite{B1975} in 1975 attracted some attention. \ However, the
introduction of zero-point energy alone did not explain many phenomena which
seemed amenable to quantum treatment. \ Many physicists who were initially
enthusiastic about the idea of zero-point energy lost interest, and some
turned actively against it. \ A careful numerical hydrogen simulation by
Cole\cite{Cole2003} introduced a rare bright spot during this situation. \ 

Apparently, it is not sufficient to introduce the idea of classical zero-point
radiation and to go on using nonrelativistic potentials for systems. \ One
must restrict attention to \textit{relativistic} or \textit{approximately
relativistic} systems. \ Also, Cole carried out numerical
calculations\cite{Cole2} for hydrogen in the presence of a single circularly
polarized plane wave propagating perpendicular to the orbit, and he noted the
presence of resonances at multiples of the orbital frequency. \ In the present
work, we find the need to consider the \textit{resonances} between the
relativistic classical charged particle and the classical zero-point
radiation. \ When we have these three ingredients, \textit{classical
zero-point radiation, relativity, and resonance}, then classical
electrodynamics produces the same results for hydrogen that Bohr's original
old quantum theory proposed. \

March 13, 2026 \ \ \ \ \ \ \ \ Hydrogen3.tex


\begin{thebibliography}{99}                                                                                               %


\bibitem {Born52}M. Born, \textit{The Mechanics of the Atom} (Ungar, New York
1970), pp. 52--53.

\bibitem {Sommerfeld}A. Sommerfeld, \textquotedblleft Zur Quantentheorie der
Spektrallinien,\textquotedblright\ Annalen der Physik, \textbf{356}, 1--94 (1916).

\bibitem {YM}W. Yourgrau and S. Mandelstam, \textit{Variation Principles in
Dynamics and Quantum Theory 3rd ed} (W. B. Saunders Co. Philadelphia, 1968),
p. 113.

\bibitem {Biedenharn}L. C. Biedenharn, \textquotedblleft The `Sommerfeld
Puzzle' Revisited and Resolved,\textquotedblright\ Found. Phys. \textbf{13},
258-279 (1983), p. 259. \ We are following Biedenharn, but our notation is
somewhat different.

\bibitem {Brev}T. H. Boyer, \textquotedblleft Stochastic Electrodynamics: The
Closest Classical Approximation to Quantum Theory,\textquotedblright\ Atoms
\textbf{7}(1), 29-38 (2019).

\bibitem {Cole2003}D. C. Cole and Y. Zou, \textquotedblleft Quantum Mechanical
Ground State of Hydrogen Obtained from Classical
Electrodynamics,\textquotedblright\ Phys. Lett. A \textbf{317}, 14-20 (2003).

\bibitem {B1975}T. H. Boyer, \textquotedblleft Random electrodynamics: The
theory of classical electrodynamics with classical electromagnetic zero-point
radiation,\textquotedblright\ Phys. Rev. D \textbf{11}, 790-808 (1975).

\bibitem {Bohr1913}N. Bohr, \textquotedblleft On the Constitution of Atom and
Molecules,\textquotedblright\ Philos. Mag. \textbf{26}, 1-25 (1913). \ 

\bibitem {SHOarticle}T. H. Boyer, \textquotedblleft The Classical Linear
Oscillator in Classical Electrodynamics with Classical Zero-Point
Radiation,\textquotedblright\ to be submitted for publication.

\bibitem {ModernTexts}See for example, R. Eisberg and R. Resnick,
\textit{Quantum Physics of Atoms, Molecules, Solids, Nuclei, and Particles
}2nd ed. (Wiley, New York 1985); K. S. Krane, \textit{Modern Physics}, 2nd ed.
(Wiley, New York 1996); R. Taylor, C. D. Zafiratos, and M. A. Dubson,
\textit{Modern Physics for Scientists and Engineers}, 2nd ed. (Pearson, New
York, 2003); S. T. Thornton and A. Rex, \textit{Modern Physics for Scientists
and Engineers}, 4th ed. (Brooks/Cole, Cengage Learning, Boston, MA, 2013).

\bibitem {Biedenharn2}See ref. 4., p260.

\bibitem {incompat}Incompatibility Between Relativistic and Nonrelativistic
Theories. \ In the \textit{nonrelativistic} limit where the speed of light $c$
is much larger than any other parameter and one subtracts off the\textit{ rest
energy} $M_{0}c^{2}$ (since $c$ is so large), the related ratio becomes a
constant,%
\begin{align*}
&  \frac{J_{\phi}\omega_{\phi}}{U-M_{0}c^{2}}\\
&  =\frac{J_{\phi}}{\left[  M_{0}c^{2}\sqrt{1-\left[  Ze^{2}/\left(  J_{\phi
}c\right)  \right]  ^{2}}-M_{0}c^{2}\right]  }\left(  \frac{M_{0}c^{3}}%
{Ze^{2}}\right)  \frac{1}{\sqrt{1-\left[  Ze^{2}/\left(  J_{\phi}c\right)
\right]  ^{2}}}\left(  \frac{Ze^{2}}{J_{\phi}c}\right)  ^{3}\\
&  \rightarrow\frac{1}{1/2\left\{  -\left[  Ze^{2}/\left(  J_{\phi}c\right)
\right]  ^{2}\right\}  }\left(  \frac{Ze^{2}}{J_{\phi}c}\right)  ^{2}=-2.
\end{align*}
Such a \textit{constant} value for $J_{\phi}\omega_{\phi}/\left[  U-M_{0}%
c^{2}\right]  $ is inconsistent with the spherical multipole radiation fields
discussed in Jackson's text and leading to the connection $\left(
dL_{z}/dr\right)  /\left(  dUdr\right)  =m/\omega$ between the radiation
angular momentum $L_{z}$ and the radiation energy $U$ in a spherical volume.

\bibitem {Casimir}H. B. G. Casimir, \textquotedblleft On the attraction
between two perfectly conducting plates,\textquotedblright\ Proc. Ned. Akad.
Wetenschap. \textbf{51}, 793-795 (1948).

\bibitem {Casimir2}M. J. Sparnaay, \textquotedblleft Measurement of the
attractive forces between flat plates,\textquotedblright\ Physica (Amsterdam)
\textbf{24}, 751-764 (1958); S. K. Lamoreaux, \textquotedblleft Demonstration
of the Casimir force in the 0.6 to 6 $\mu$m range,\textquotedblright\ Phys.
Rev. Lett. \textbf{78}, 5-8 (1997): \textbf{81}, 5475-5476 (1998); U.
Mohideen, \textquotedblleft Precision measurement of the Casimir force from
0.1 to 0.9 $\mu$m,\textquotedblright\ \textit{ibid.} \textbf{81}, 4549-4552
(1998); H. B. Chan, V. A. Aksyuk, R. N. Kleinman, D. J. Bishop, and F.
Capasso, \textquotedblleft Quantum mechanical actuation of
microelectromechanical systems by the Casimir force,\textquotedblright%
\ Science \textbf{291}, 1941-1944 (2001): G. Bressi, G. Carugno, R. Onofrio,
and G. Ruoso, \textquotedblleft Measurement of the Casimir force between
parallel metallic surfaces,\textquotedblright\ Phys. Rev. Lett. \textbf{88},
041804(4) (2002). \ 

\bibitem {Marshall}T. W. Marshall, \textquotedblleft Statistical
Electrodynamics,\textquotedblright\ Proc. Camb. Phil. Soc. \textbf{61},
537-546 (1965).

\bibitem {Jackson}J. D. Jackson, \textit{Classical Electrodynamics 2nd ed}
(John Wiley \& Sons, New York, 1975), p. 746.

\bibitem {EH}A. Einstein and L. Hopf, \textquotedblleft\"{U}ber einen Satz der
Wahrscheinlichkeitsrechnung und seine Anwendung in der
Strahlungstheorie,\textquotedblright\ Annalen der Physik (Leipzig)
\textbf{33}, 1096-1104 (1910).

\bibitem {Rice}Discussion of random radiation in terms of random phases can be
found in the article by S. O. Rice, \textquotedblleft Mathematical Analysis of
Random Noise,\textquotedblright\ in \textit{Selected papers on Noise and
Stochastic Processes}, edited by N. Wax (Dover, New York 1954) , p. 138. \ 

\bibitem {Jackson2}See ref. 16, p. 746.

\bibitem {Jackson3}See ref. 16, pp. 99-100.

\bibitem {Disguised}T. H. Boyer, \textquotedblleft Disguised electromagnetic
connections in classical electron theory,\textquotedblright\ Eur. J. Phys.
\textbf{43, }025201 (21pp) (2022).

\bibitem {Born2}M. Born, \textit{Atomic Physics 7th ed.} (Hafner, New York
1966), pp. 116-117.

\bibitem {Greenberg}M. D. Greenberg, \textquotedblleft\textit{Advanced
Engineering Mathematics},2nd. ed\textquotedblright\ (Prentice Hall, Upper
Saddle River, NJ, 1998), pp. 141-144.

\bibitem {Burko}L. M. Burko, \textquotedblleft Self-force approach to
synchrotron radiation,\textquotedblright\ Am. J. Phys. \textbf{68}, 456-468
(2000). \ See p. 463, Eq. (46).

\bibitem {Jackson4}See ref. 16, Problem 9.1, pp. 460-461.

\bibitem {Nernst}W. Nernst, \textquotedblleft\"{U}ber einen Versuch, von
quantentheoretischen Betrachtungen zur Annahme stetiger Energie\"{a}nderungen
zur\"{u}ckzukehrn,\textquotedblright\ Verhandlungen der Deutschen
Physikalischen Gesellschaft \textbf{18}, 83--116 (1916).

\bibitem {Marshall2}T. W. Marshall, \textquotedblleft Random
electrodynamics,\textquotedblright\ Proc. R. Soc. \textbf{A276}, 475-491 (1963).

\bibitem {Cole2}D. C. Cole, \textquotedblleft Subharmonic resonance and
critical eccentricity for the classical hydrogen atomic
system,\textquotedblright\ Eur. Phys. J. D \textbf{72}, 200-214 (2018). \ D.
C. Cole and Y. Zou, \textquotedblleft Subharmonic resonance behavior for the
classcal hydrogen atomic system,\textquotedblright\ J. Sci. Comput. 39, 1-27 (2009).
\end{thebibliography}
\end{document}